\documentclass[english]{article}
\pdfoutput=1
\usepackage[T1]{fontenc}
\usepackage[latin9]{inputenc}
\usepackage{geometry}
\usepackage{units}
\usepackage{amstext}
\usepackage{amssymb}
\usepackage{stmaryrd}
\usepackage{graphicx}
\usepackage{esint}



\usepackage{babel}
\begin{document}

\title{Thermodynamic properties of diatomic molecules systems under anharmonic
Eckart potential}

\author{G. Valencia--Ortega\textsuperscript{a} and L. A. Arias--Hernandez\textsuperscript{b}\\
 Departamento de Física, Escuela Superior de Física y Matemáticas,
Instituto\\
 Politécnico Nacional, U. P. Zacatenco, edif. \#9, 2o Piso, Ciudad
de México,\\
 07738, MÉXICO, gvalencia@esfm.ipn.mx\textsuperscript{a} and larias@esfm.ipn.mx\textsuperscript{b}}
\maketitle

\begin{abstract}
Due to one of the most representative contributions to the energy
in diatomic molecules being the vibrational, we consider the generalized
Morse potential (GMP) as one of the typical potential of interaction
for one\textendash dimensional microscopic systems, which describes
local anharmonic effects. From Eckart potential (EP) model, it is
possible to find a connection with the GMP model, as well as obtain
the analytical expression for the energy spectrum because it is based
on $S\,O\left(2,1\right)$ algebras. In this work we find the macroscopic
properties such as vibrational mean energy $U$, specific heat $C$,
Helmholtz free energy $F$ and entropy $S$ for a heteronuclear diatomic
system, along with the exact partition function and its approximation
for the high temperature region. Finally, we make a comparison between
the graphs of some thermodynamic functions obtained with the GMP and
the Morse potential (MP) for $H\,Cl$ molecules.

\vspace{0.3cm}
Keywords: generalized Morse potential, thermodynamic functions.{\small \par}
\end{abstract}

\section{Introduction}

In the study of systems made up of many particles and phenomena of
matter, statistical physics has been able to interpret and predict
several macroscopic properties from them through the average of dynamic
amounts over a specific number of particles. To obtain the macroscopic
variables it is necessary to take into account that for real macroscopic
systems the transitions between their quantum states occur randomly
and quickly; for this reason, these macroscopic observables depend
on the average of the expected dynamic values \cite{GreinerNeiseStoecker95,Callen85}.
From the point of view of quantum statistical mechanics it is particularly
interesting that to resolve the thermodynamics of these systems, we
must count with their dynamic equation (Eq. Schrödinger) which contains
as close as possible the real information on the evolution of all
above systems. A detailed study for any molecule should consider all
the contributions to the energy, because in most cases one establishes
that intrinsic movements (vibrations, rotations, etc.) of them that
shape a particular gas (system) are independent of each other, then
we proceed to study them separately \cite{McQuarrie76,Levine01}.

The experiments based on spectroscopy show that there are anharmonicities
in real molecular vibrations systems \cite{Levine01,Herzberg50},
therefore it is essential to know the algebraic model for the potential
and the technique used to solve the Schrödinger equation. For example,
anharmonicities are usually introduced in some models as a perturbation
to the harmonic oscillator \cite{Dong07,DongLozadaYuJimenezRivera07}.
There are several potential that can be considered to emulate the
molecular vibrational spectrum of a gas composed of diatomic molecules
(section 2) and can be compared through the experimental data \cite{StaffordHoltPaulson63}.
The purpose of this work is to compare the thermodynamic properties
of two anharmonic quantum models for one-dimensional systems (1D)
\cite{DengFan57}, which consist of exact solutions.

The fundamental characteristic of these anharmonic potentials consist
of a finite energy spectrum and anharmonicities are introduced by
means of dynamic groups obeying a particular algebra. Using the connection
between the dynamic parameters of the potential GMP and EP, it is
possible to take a more manageable expression to obtain the eigenvalues
\cite{CodrianskyCorderoSalam99}, such that systems under study consist
of heteronuclear molecules formed by a hydrogen ion and one of the
four most electronegative ions of the VII A group in the periodic
table ($H\,F$, $H\,Cl$, $H\,Br$ and $H\,I$ ), we take into account
also that the molecules are formed in the ground electronic state.
In Section 3, we derive the partition function without an approximation
i.e. the ``Zustandssumme'' and with the approximation of high vibrational
temperatures to obtain the basic thermodynamic functions, such as
the mean vibrational energy, specific heat, free energy, etc. since
they are only in terms of the model parameters. In Section 4 a comparative
analysis of some thermodynamic functions is obtained by the GMP and
the MP, that has been studied in recent works and are applied to $H\,Cl$
system \cite{AngelovaFrank05}. Eventually, the concluding remarks
are given in Section 5.

\section{Generalized Morse Hamiltonian model}

We considered initially the Schrödinger equation (1D) whose interaction
is mediated through a potential introduced by Deng and Fan \cite{DengFan57}.
This potential describes the energy spectrum of diatomic molecules
as well as their electromagnetic transitions, and is refered to literature
as the generalized Morse potential (GMP) \cite{CodrianskyCorderoSalam99,DelsolQuesneSmirnov98},

\begin{equation}
\left[-\frac{\hbar^{2}}{2\mu}\frac{\partial^{2}}{\partial r^{2}}+D\left(\frac{b}{\exp\left(ar\right)-1}\right)\right]\psi(r)=E\psi(r);\:\:\:b=\exp\left(ar_{e}\right)-1.\label{potgmp}
\end{equation}

Where $0\leq r<\infty$, $D$ is the depth of the potential well,
$a$ is related with its width, $r_{e}$ is the equilibrium position
with respect to a given origin and $\mu$ is the reduced mass of the
molecule. Sometimes the GMP is related to the Manning\textendash Rose
potential also known as Eckart potential (EP). One can verify that
this potential has the same behavior as the internuclear potential
of diatomic molecules \cite{Herzberg50,DelsolQuesneSmirnov98}.

In order to obtain a dynamic equation with dimensionless variables,
we make the change of variable $x=ar$, the Schrödinger equation (\ref{potgmp})
is rewritten as:

\begin{equation}
\left[-\frac{d^{2}}{dx^{2}}+\frac{2\mu D}{a^{2}\hbar^{2}}\left(1-\frac{b}{\exp x-1}\right)^{2}\right]\psi(x)=\frac{2\mu E}{a^{2}\hbar^{2}}\psi(x),\label{potgmp2}
\end{equation}
there are several ways to solve these kind of equations, one of them
is the analytical form presented in \cite{DengFan57}, in which by
means of a new variable $y$ and a function $F(y)$, one can define

\begin{equation}
y=\left(\exp x-1\right)^{-1}\label{newy}
\end{equation}

\begin{equation}
\psi(x)=\Phi(y)=\frac{y^{\eta}}{\left(1+y\right)^{\nu}}F(y),\label{newFy}
\end{equation}
where $\alpha$ and $\beta$ are chosen in a proper way, so there
is a physical sense. When introducing (\ref{newy}) and (\ref{newFy})
into (\ref{potgmp2}), we obtain the following equation:

\begin{equation}
y\left(1+y\right)\frac{d^{2}F}{dy^{2}}+\left[2\left(\eta-\nu+1\right)y+\left(2\eta+1\right)\right]\frac{dF}{dy}+\left[\left(\eta-\nu\right)^{2}+\left(\eta-\nu\right)-\frac{2\mu D}{a^{2}\hbar^{2}}b^{2}\right]F(y)=0.\label{newpotgmp}
\end{equation}
One of the solutions already explored, corresponds to a confluent
hypergeometric function and is given as follows: $F(y)=_{2}F_{1}\left(d;\,e;\,2\eta+1;\,-y\right)$
with

\begin{equation}
\begin{array}{c}
d=\eta-\nu+l\\
e=\eta-\nu+1-l\\
l\equiv\frac{1}{2}\left(1+\sqrt{1+kb^{2}}\right)=\frac{1}{2}\left(1+\sqrt{1+\frac{2\mu D}{a^{2}\hbar^{2}}b^{2}}\right)
\end{array}\label{defparam}
\end{equation}

As it is desired that the $\eta$ parameter is related to the system
energy spectrum (solution of Eq. \ref{newpotgmp}), then each $\epsilon_{n}$
can be expressed in terms of a $\eta_{n}$, such that if $\eta=\sqrt{k-\epsilon}\;\Longrightarrow\;\epsilon_{n}=k-\eta_{n}^{2}$,
with $k$ a fixed parameter so the energy spectrum becomes finite.
It is concluded that the energy eigenvalues are given by:

\begin{equation}
E_{n}=D-\frac{a^{2}\hbar^{2}}{8\mu}\left(n+l-\frac{bk\left(b+2\right)}{n+l}\right).^{2}\label{enerespecgmp}
\end{equation}

It is important to note that the quantum numbers $n$ form a finite
set of values, whose supreme $n_{max}$ is assumed integer (for the
real case of molecular interactions this value has to be taken as
its integer part),

\begin{equation}
n=0,\,1,\,2,\,\ldots\,,\,n_{max};\:\:n_{max}=\sqrt{kb\left(b+2\right)}-l.\label{valmax}
\end{equation}

\subsection{Connection between GMP and EP potentials}

It has been found that in the eigenvalue problem there are symmetry
algebras that offer an alternative solution to those already well\textendash known.
In particular a set of techniques based on the $S\,O\left(2,\,1\right)$
algebra can be used, whose purpose is to build the hypergeometric
Natanzon potentials \cite{CodrianskyCorderoSalam99,Natanzon79}. The
aim of the above is to transform a Schrödinger equation in one of
these potentials through $S\,O\left(2,\,1\right)$ algebra, taking
into account that their generator groups $J_{0}$ and $J_{\pm}=J_{1}\pm\textrm{i}J_{2}$
form the Casimir operator $C$ that is written as $C=J_{0}\left(J_{0}\pm1\right)-J_{\mp}J_{\pm}$.
The Schrödinger equation is rewritten in terms of the Casimir operator
and takes the form $\left[H-E\right]\Psi(r,\,\phi)=G(r)\left[C-q\right]\Psi(r,\,\phi)$,
where $q$ is the eigenvalue of $C$ whereas $E$ is the corresponding
eigenvalue of the Hamiltonian $H$.

In general, GMP and EP are given by

\begin{equation}
V_{GMP}=K_{1}\left(1-\frac{K_{2}}{\exp(\omega r)-1}\right)+K_{3},\label{gmp}
\end{equation}

\begin{equation}
V_{E}=A^{2}+\frac{B^{2}}{A^{2}}-2B\coth(\alpha r)+A\left(A-\alpha\right)\textrm{csch}(\alpha r)^{2}\label{ep}
\end{equation}
due to the set of constants $\left(K_{1},\,K_{2},\,K_{3}\right)$
and $\left(A,\,B,\,\alpha\right)$ can be related as follows:

\begin{equation}
\begin{array}{c}
A^{2}+\frac{B^{2}}{A^{2}}=\left(1+K_{2}+\frac{1}{2}K_{2}^{2}\right)K_{1}+K_{3};\;\;\;2B=\frac{1}{2}K_{1}K_{2}\left(K_{2}+2\right);\\
A\left(A-\alpha\right)=\frac{1}{4}K_{1}K_{2}^{2};\;\;\;\alpha=\frac{\omega}{2}
\end{array}.\label{eqvconst}
\end{equation}

The $V_{GMP}$ and $V_{E}$ are the same function. The functional
form of the hypergeometric Natanzon potentials are given by \cite{Natanzon79,CooperGinocchioKhare87}

\begin{equation}
V_{N}=\frac{1}{R}\left[f\,z\left(r\right)^{2}-\left(h_{0}-h_{1}+f\right)z\left(r\right)+h_{0}+1\right]+\frac{z\left(r\right)^{2}\left(1-z\left(r\right)\right)^{2}}{R^{2}}\left[a+\frac{a+\left(c_{1}-c_{0}\right)\left(2\,Z\left(r\right)-1\right)}{z\left(r\right)\left(z\left(r\right)-1\right)}-\frac{5\triangle}{4R}\right],\label{potnatan}
\end{equation}
with $\triangle=\tau^{2}-4ac_{0}$, $\tau=c_{1}-c_{0}-a$ and $R=az(r)^{2}+\tau z(r)+c_{0}$.
The constants $\left(a,\,c_{0},\,c_{1},\,h_{0},\,h_{1},\,f\right)$
are called Natanzon parameters, besides the function $z(r)$ has to
satisfy

\begin{equation}
\frac{dz(r)}{dr}=\frac{2z(r)\left(1-z(r)\right)}{\sqrt{R}}.\label{condzr}
\end{equation}

One can check that the Natanzon parameters for Eckart potential are

\begin{equation}
\begin{array}{c}
a=c_{0}=\alpha^{-2}\;\;\;c_{1}=0\;\;\;h_{0}=\frac{\left(A^{2}+B\right)^{2}}{A^{2}\alpha^{2}}-1\\
h_{1}=\frac{4A\left(A-\alpha\right)}{\alpha^{2}}\;\;\;f=\frac{\left(A^{2}-B\right)^{2}}{A^{2}\alpha^{2}}-1
\end{array}\label{nataparam}
\end{equation}
and $z(r)$ must be

\begin{equation}
z(r)=\exp\left(2\alpha r\right).\label{zr}
\end{equation}

With all the above, the energy spectrum of EP can be obtained according
to Codriansky et al \cite{CodrianskyCorderoSalam99}, where it is
required that $E_{(n=0)}=0$ and this result leads to

\begin{equation}
E_{n}^{GMP}=A^{2}+\frac{B^{2}}{A^{2}}-\left(A+\alpha n\right)^{2}-\frac{B^{2}}{\left(A+\alpha n\right)^{2}},\:\:\:n=0,\,\ldots\,,\,n_{max}\label{enerespecep}
\end{equation}
the value $n_{max}$ corresponds to an upper bound of $E_{n}$, which
accomplishes $E(n)\leq V_{E}(r\rightarrow\infty)=\nicefrac{\left(B-A^{2}\right)^{2}}{A^{2}}$
and therefore from Eq. (\ref{enerespecep}) we obtain $n_{max}=\left\llbracket \nicefrac{\left(\sqrt{B}-A\right)}{\alpha}\right\rrbracket $,
where $\left\llbracket \right\rrbracket $ means the integer part
of $\nicefrac{\left(\sqrt{B}-A\right)}{\alpha}$, and also $B>A^{2}$
has to be accomplished.

If we compare the expression obtained for the GMP energy spectrum
(\ref{enerespecgmp}) and the other deduced for the EP (\ref{enerespecep}),
we note that this last expression is more useful for the following.

\section{Macroscopic response of diatomic molecular systems}

One of the objectives of thermodynamics consists of estimating the
free energy (what can be converted in work) which is obtained from
the interchanged heat and mass by the system under study and the surroundings.
This macroscopic behavior is the result of interactions that occur
between the constituent elements of each system. In general they can
be considered open or closed systems. In order to find an expression
that describes the behavior of these systems that consider both the
restrictions imposed and its evolution to the final equilibrium state,
by means of the ensemble theory, we can build from the averages of
microscopic quantities the fundamental equation of such systems \cite{Pathria96}.

Some of the systems whose microscopic components can factor their
internal freedom degrees due to its well-defined symmetry are diatomic
molecules. In particular we consider the set of gases made\textendash up
of heteronuclear molecules: $H\,F$,$H\,Cl$,$H\,Br$ and $H\,I$,
for this set the contribution to the total energy with more statistical
weight corresponds to the vibrational energy \cite{McQuarrie76,Herzberg50}.
Naturally, it provides that each microstate of the system is characterized
by extensive set of variables $\left(N,\,V,\,T\right)$ with energy
$E_{n}$.

\subsection{Diatomic molecular system at the canonical ensemble for the GMP}

Before studying the thermodynamic properties of the above set of molecules,
it is necessary to obtain the anharmonic vibrational individual partition
function of this kind of molecules:

\begin{equation}
z^{GMP}=\sum_{n=0}^{\left\llbracket \nicefrac{\sqrt{B}-A}{\alpha}\right\rrbracket }e^{-\beta E_{n}},\label{partfunct}
\end{equation}
where $\beta=\nicefrac{1}{k_{B}T}$, whereas the anharmonic energy
spectrum $E_{n}$ is given by (\ref{enerespecep}). The individual
partition function is determined by the parameters $\left(A,\,B,\,\alpha\right)$,
which can be directly related to the mechanical energy of molecules.
Introducing the parameter $P=A^{2}+\nicefrac{B^{2}}{A^{2}}$, the
exact partition function can be written as:

\begin{equation}
z^{GMP}=e^{-\beta P}\sum_{n=0}^{\left\llbracket \nicefrac{\sqrt{B}-A}{\alpha}\right\rrbracket }\exp\beta\left[\left(A+\alpha n\right)^{2}+\frac{B^{2}}{\left(A+\alpha n\right)^{2}}\right].\label{partfunctexac}
\end{equation}

When one considers the high temperature region $T\gg1$ the energy
states can be seen almost as a continuous spectrum, therefore we can
replace with good approximation the sum by the integral in (Eq. \ref{partfunctexac}):

\begin{equation}
\tilde{z}^{GMP}=\frac{e^{-\beta P}}{\alpha}\int_{\left\llbracket A\right\rrbracket }^{\left\llbracket \sqrt{B}\right\rrbracket }\exp\left[\beta\left(m^{2}+\frac{B^{2}}{m^{2}}\right)\right]dm,\label{partfunctint}
\end{equation}
with $m=A+\alpha n$. This integral can be evaluated exactly using
the properties of the denote integral as well as the imaginary error
function $\textrm{erfi}(x)$,

\begin{equation}
\tilde{z}^{GMP}=\frac{e^{-\beta P}}{\alpha}\left[\frac{e^{-2\beta B}\sqrt{\pi}}{4\sqrt{\beta}}\left(e^{4\beta B}\textrm{erfi}\left[\left(\frac{B}{A}-A\right)\sqrt{\beta}\right]+\textrm{erfi}\left[2\sqrt{\beta B}\right]-\textrm{erfi}\left[\left(\frac{B}{A}+A\right)\sqrt{\beta}\right]\right)\right].\label{partfunctappr}
\end{equation}
\begin{table}
\begin{center}
\begin{tabular}{|c|c|c|c|c|}\hline
\textbf{Molecule} & $\mathbf{\mu}$ \textrm{(Kg)} & $D$ \textrm{(J)} & $d$ \textrm{(m)} & $r_{e}$ \textrm{(m)} \\ \hline
  $H\,F$ & $1.5897\times10^{-27}$ & $9.430\times10^{-19}$ & $0.430\times10^{-10}$ & $0.9171\times10^{-10}$ \\ \hline
  $H\,Cl$ & $1.6276\times10^{-27}$ & $8.520\times10^{-19}$ & $0.570\times10^{-10}$ & $1.2746\times10^{-10}$ \\ \hline
  $H\,Br$ & $1.6531\times10^{-27}$ & $7.705\times10^{-19}$ & $0.605\times10^{-10}$ & $1.4130\times10^{-10}$ \\ \hline
  $H\,I$ & $1.6608\times10^{-27}$ & $6.648\times10^{-19}$ & $0.650\times10^{-10}$ & $1.6040\times10^{-10}$ \\ \hline
\end{tabular}
\end{center}
\caption{\label{spectdiatmole} Spectroscopic data of heteronuclear molecules
formed by a hydrogen ion and ions of the VII A group}
\end{table}
This equation represents the extended sum over all accessible microstates
in the high temperatures region whose interaction is of the type GMP,
and is only a function of parameters which are directly related to
the energy that the system can exchange. For the set of diatomic molecules
considered, it is necessary to list some useful values for their energetic
study. In Table \ref{spectdiatmole} we can find the reduced mass $\mu$, the depths of of the GMP well (dissociation
energies) $D$, the widths of GMP well $d$ and the equilibrium points
of the interaction potential $r_{e}$. These values can be derived
directly or indirectly from the spectroscopic data of the emission
lines and we specially focus our attention in the $X^{1}\Sigma^{+}$
state for each molecule \cite{Herzberg50}.

By solving simultaneously (\ref{eqvconst}), we find the values of
$A$ and $B$ as follows:

\begin{equation}
\begin{array}{c}
B=\frac{1}{4}Db\left(b+2\right),\\
A^{2}+\frac{B^{2}}{A^{2}}=K_{1}\;\Rightarrow\;A^{2}=\frac{K_{1}-\sqrt{K_{1}-4B^{2}}}{2},\;\;\;\;K_{1}=D\left(1+b+\frac{1}{2}b^{2}\right).
\end{array}\label{constmatch}
\end{equation}

\begin{figure}
\begin{centering}
\includegraphics{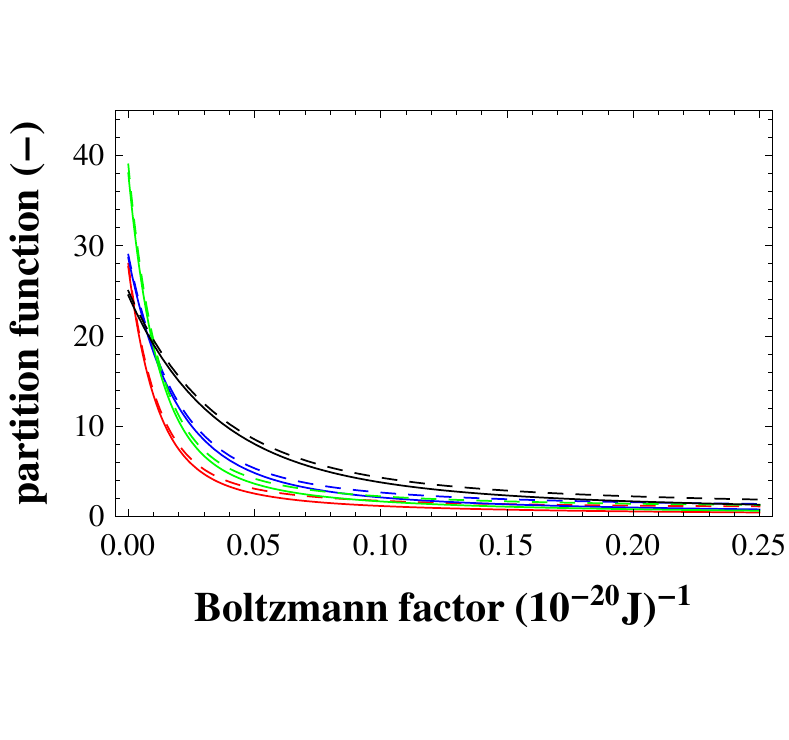}
\par\end{centering}
\caption{\label{figpartfunct}Graphs of anharmonic vibrational individual partition
functions $z^{GMP}$ (solid lines) and $\tilde{z}^{GMP}$ (dashed
lines) of $H\,F$ (red), $H\,Cl$ (blue), $H\,Br$ (green) and $H\,I$
(black) as functions of $\beta$. Solid lines represent the states
that are counted in the region of high temperatures, while dashed
lines denote all states that can be in all temperatures.}
\end{figure}

Substituting the values of $\left\llbracket A\right\rrbracket $ and
$\left\llbracket \sqrt{B}\right\rrbracket $ for each molecule, we
can obtain numerically the individual partition function $\tilde{z}^{GMP}$
as a function of $\beta$. In Fig. \ref{figpartfunct} the approximate
partition functions (solid curve) for values of $\beta$ between $0$
and $0.25$ are presented, as well as the exact partition functions
(dashed curve).
 As we can observe in the region $0<\beta<0.25$, $\tilde{z}^{GMP}$
is a good approximation of the exact ``Zustandssumme'' $z^{GMP}$.
In this interval one can verify that the integrals are slightly offset
from value with respect to its accurate representation compared from
their relative minimum. On the other hand, in the region corresponding
to $0.25<\beta<1$, we can verify the eight lines (approximate and
exact) overlap. The statistical physics which follows from this model
is a first approximation; for the realizations and transitions between
quantum states of molecules, this approximation is valid for the temperature
region lower than its dissociation energy. Nevertheless, a more realistic
description requires states not tied to the potential.

Once we have found the individual partition function $\tilde{z}^{GMP}$
in terms of GMP parameters, we can build the corresponding thermodynamic
functions, which are presented in the following subsection.

\subsubsection{Thermodynamic functions}

Now we will study the thermodynamic properties of these systems derived
from canonical ensemble.

The molar vibrational internal energy in the region of high temperature,
is given by:

\begin{equation}
u^{GMP}=-\frac{\partial}{\partial\beta}\ln\tilde{z}^{GMP}=-\frac{1}{\tilde{z}^{GMP}}\frac{\partial\tilde{z}^{GMP}}{\partial\beta}.\label{intenerg}
\end{equation}

\begin{figure}
\begin{centering}
\includegraphics{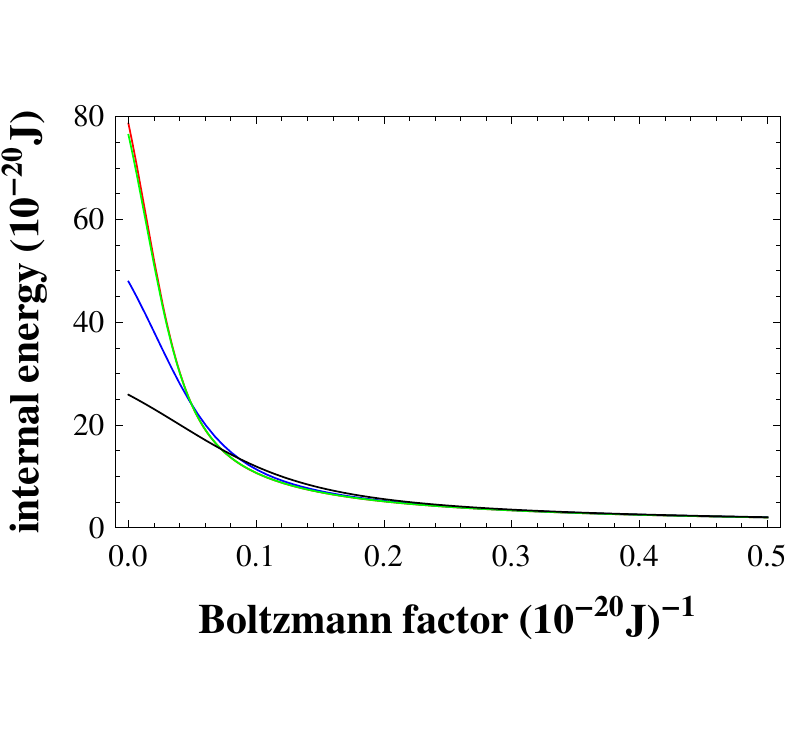}
\par\end{centering}
\caption{\label{figintenerg}Graphs of anharmonic vibrational individual internal
energy $u$ of $H\,F$ (red line), $H\,Cl$ (blue line), $H\,Br$
(green line) and $H\,I$ (black line) as function of $\beta$.}
\end{figure}

If the analytical expression of $z^{GMP}$ (Eq \ref{partfunctappr})
is taken and then is replaced in (\ref{intenerg}), an analytical
expression for $u$ is obtained. It is clear that $u$ depends on
the parameters $\left(A,\,B,\,\alpha\right)$. Figure \ref{figintenerg}
shows the average vibrational energy of each molecule that has already
been mentioned above 
. We can see in the region $\left(0<\beta<0.5\right)$, the curves
exhibit the well\textendash known characteristic monotone decreasing
behavior. In $\beta=0$, the value of $u$ of the 4 systems correspond
to different values. This is because of the GMP depth generated by
each molecule is different and therefore the partition function counts
different states, whereas in the range $\left(0.5<\beta<1\right)$
curves are overlaped.

Now we consider the vibrational contribution to the Helmholtz free
energy per molecule in terms of the partition function $\tilde{z}$
which is calculated for the region of high temperatures as

\begin{equation}
\begin{array}{c}
f^{GMP}=-\frac{1}{\beta}\ln\tilde{z}^{GMP}\\
=-\frac{1}{\beta}\ln\left\{ \frac{e^{-\beta K_{1}}}{\alpha}\left[\frac{e^{-2\beta B}\sqrt{\pi}}{4\sqrt{\beta}}\left(e^{4\beta B}\textrm{erfi}\left[\left(\frac{B}{A}-A\right)\sqrt{\beta}\right]\right)+\textrm{erfi}\left[2\sqrt{\beta B}\right]-\textrm{erfi}\left[\left(\frac{B}{A}+A\right)\sqrt{\beta}\right]\right]\right\} 
\end{array}.\label{freeenerg}
\end{equation}

\begin{figure}
\begin{centering}
\includegraphics{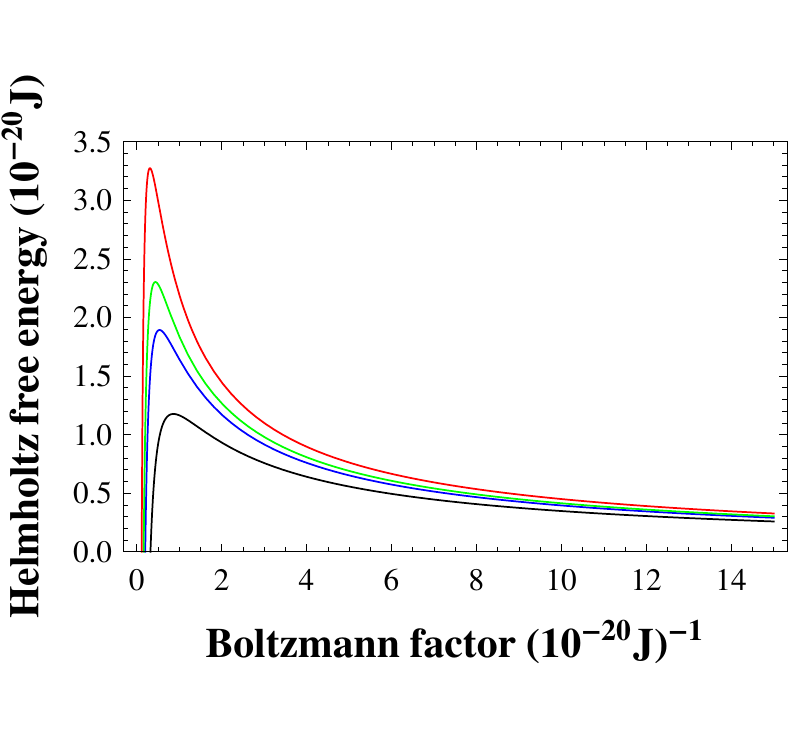}
\par\end{centering}
\caption{\label{figfreeenerg}Anharmonic vibrational contribution to the free
energy per molecule $f^{GMP}$ of $H\,F$ (red line), $H\,Cl$ (blue
line) $H\,Br$ (green line) and $H\,I$ (black line) as function of
$\beta$. It is observed that the maximum of the functions will reach
larger values when the interaction potential hosts more states.}
\end{figure}

We observe in Figure \ref{figfreeenerg} that the curves of the molar
Helmholtz free energy attain maximum values for different temperatures,
which are $\beta_{HF}=0.3107$, $\beta_{HCl}=0.5366$, $\beta_{HBr}=0.4395$
and $\beta_{HI}=0.8647$
. In addition, for systems ($H\,F$, $H\,Cl$, $H\,Br$ and $H\,I$
) their free energies decay asymptotically in the region of low temperatures.
As we can see, there is a hierarchy for each system composed of one
of four different molecules which are in contact with a thermal reservoir.
So it can be interpreted as the available chemical work that sets
out every gas. The graph \ref{figfreeenerg} agrees with the experimental
fact, in which a $H\,F$\textendash gas has the corrosive properties
severer than the others.

The vibrational specific heat at constant volume is obtained from
the expression:

\begin{equation}
c_{v}^{GMP}=\frac{\partial u^{GMP}}{\partial T}=-k_{B}\beta^{2}\frac{\partial u^{GMP}}{\partial\beta}.\label{specheat}
\end{equation}

\begin{figure}
\begin{centering}
\includegraphics{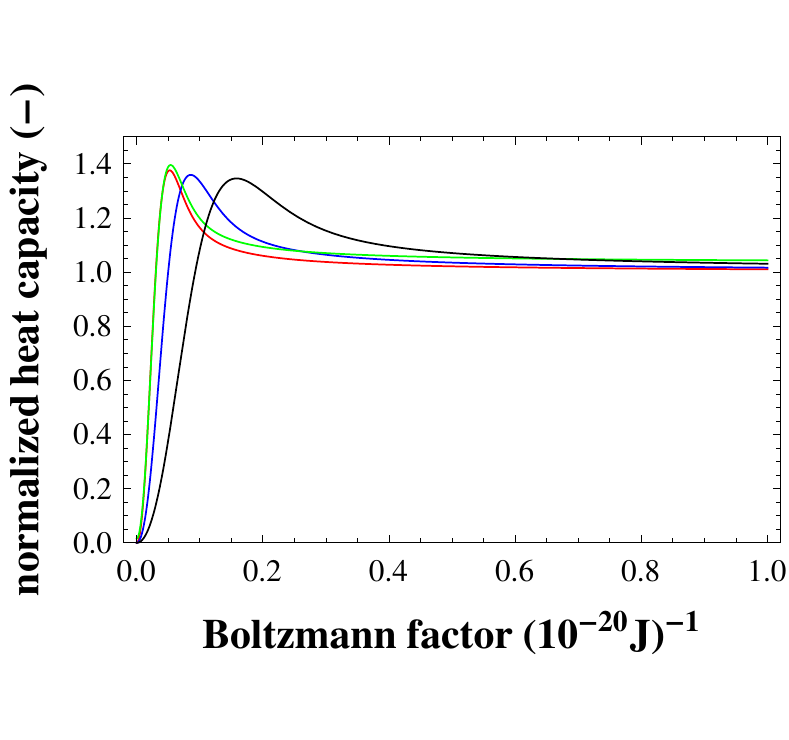}
\par\end{centering}
\caption{\label{figspecheat}Specific heat $\nicefrac{c_{v}^{GMP}}{k_{B}}$
for $H\,F$ (red line), $H\,Cl$ (blue line), $H\,Br$ (green line)
and $H\,I$ (black line) as function of $\beta$.}
\end{figure}

By substituting the expression from (\ref{intenerg}) into (\ref{specheat}),
we verify that $\nicefrac{c_{v}^{GMP}}{k_{B}}$ (normalized by the
Boltzmann constant) is also expressed in terms of the parameters of
the potential. In Figure \ref{figspecheat} we show the anomalous
behavior of the specific heat when diatomic systems interact under
the GMP
. Anharmonicity effects are apparent to the values $\beta_{HF}\leq0.120$,
$\beta_{HCl}\leq0.168$, $\beta_{HBr}\leq0.121$ and $\beta_{HI}\leq0.221$,
each of these values correspond to the characteristic vibrational
temperatures of the molecules ($\Theta_{HF}\approx5954.11\textrm{K}$,
$\Theta_{HCl}\approx4300\textrm{K}$, $\Theta_{HBr}\approx3787\textrm{K}$
and $\Theta_{HI}\approx3266\textrm{K}$) \cite{Levine01,Herzberg50}.
Anharmonics contributions of the analytical model (\ref{specheat})
depend mainly on the parameter $\beta$. The specific heat for this
model has a maximum for a value $\beta=\beta_{C}$ corresponding to
a temperature $T=T_{C}$. This is due to the finite number of states
of the algebraic model. It means for a system composed of diatomic
molecules whose potential energy is of the type GMP, it has a critical
temperature value in which the system become saturated and can no
longer absorb more energy because all its exited states are occupied.

This model illustrates the behavior of the specific heat without regard
to other degrees of freedom, so the relative value $T_{C}$, for each
molecule is $T_{C}^{HF}=13662.1\textrm{K}$, $T_{C}^{HCl}=8440.8\textrm{K}$,
$T_{C}^{HBr}=13424.8\textrm{K}$ and $T_{C}^{HI}=4570.4\textrm{K}$.
On the other hand, the region $T>T_{C}$ considers the continuum states,
therefore the most important contribution is the translational.

Finally, we can express the entropy as a state function in terms of
the partition function, the vibrational individual contribution to
the entropy is obtained as

\begin{equation}
s^{GMP}=k_{B}\ln\tilde{z}^{GMP}+k_{B}\left(\frac{\partial\ln\tilde{z}^{GMP}}{\partial T}\right)=k_{B}\left[\ln\tilde{z}^{GMP}-\frac{\beta}{\tilde{z}^{GMP}}\frac{\partial\tilde{z}^{GMP}}{\partial\beta}\right].\label{entrop}
\end{equation}

In this section we have obtained the most important thermodynamic
functions, all in terms of the imaginary function error $\textrm{erfi}(x)$,
because the energy spectrum is bounded for the GMP.

\section{A comparison of some thermodynamic properties between GMP and Morse}

\begin{figure}[tb]
\noindent \begin{centering}
\includegraphics{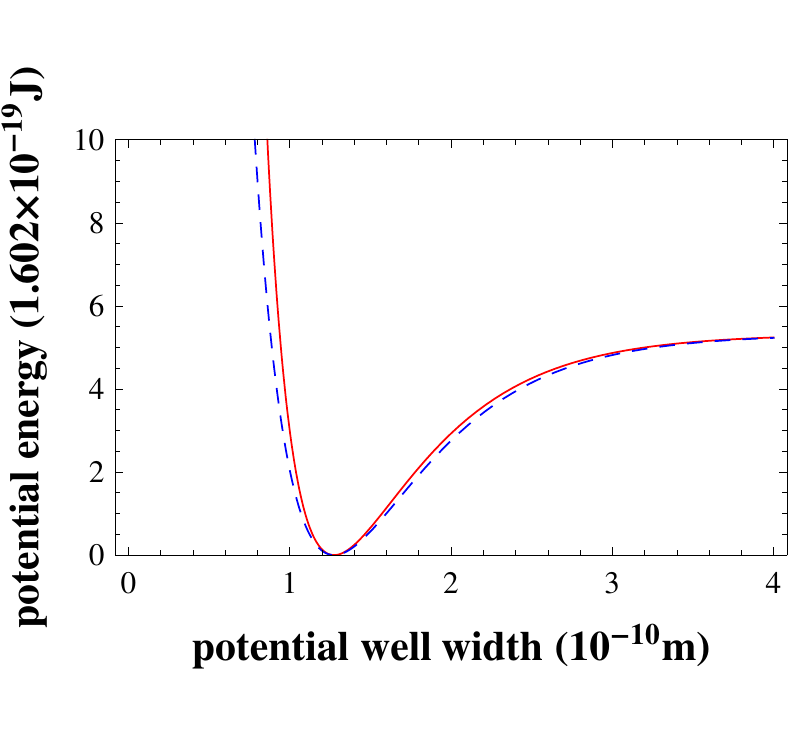} 
\par\end{centering}
\caption{\label{figcompgmpandmors.}Comparison between the curves of the GMP
(solid line) and the Morse potential (dashed line). The parameters
used have the following values $d^{-1}=a=1.7543\times10^{10}\textrm{\ensuremath{\textrm{\ensuremath{m^{-1}}}}}$,
$r_{e}\:\textrm{or}\:x_{e}=1.2746\times10^{-10}\textrm{\ensuremath{\textrm{m}}}$
y $D=8.52\times10^{-19}\,\textrm{J}$, which are the spectroscopic
values for the molecule $H\,Cl$.}
\end{figure}

There are many proposals to study the collective behavior of a system
of diatomic molecules via the algebraic model of a potential interaction
between the two ions \cite{Dong07,AngelovaFrank05}. Morse potential
has been a good proposal, since in recent articles has been resolved
exactly and therefore the vibrational thermodynamic functions of the
gas have been obtained. It is considered that the expression of one-dimensional
Hamiltonian for the MP is,

\begin{equation}
\mathcal{H}=-\frac{\hbar^{2}}{2\mu}\frac{\partial^{2}}{\partial x^{2}}+D\left(1-\exp\left[-\frac{\left(x-x_{e}\right)}{d}\right]\right)^{2},\label{morsepot}
\end{equation}
with $D$ the depth of potential, $d$ the width, $x_{e}$ the displacement
from the equilibrium position and $\mu$ the reduced mass of the oscillating
system. We note that MP is a particular case of the GMP when $b=1$
and when $\exp\left(\nicefrac{x}{d}\right)\gg1$ (see Fig. \ref{figcompgmpandmors.})
. The individual vibrational spectrum of energy is written as

\begin{equation}
E_{n}^{MP}=\hbar\omega_{0}\left(n+\frac{1}{2}-\frac{n^{2}}{N}\right),\;\;\;\;n=0,\ldots,\frac{N}{2},\label{mpenerspec}
\end{equation}
where $\omega_{0}$ is the harmonic oscillator frequency and $n$
is the number of quanta in the oscillator. M. Angelova et al \cite{AngelovaFrank05}
obtained analytical expressions of some thermodynamic vibrational
functions. Some worth remarking with regard to this potential model
is the dependency of the $N_{0}$ parameter, which is the maximum
number of bosons per oscillator. Thus, the individual partition function,
molar internal energy and specific heat for the MP are as follow,

\begin{equation}
\begin{array}{c}
z_{N}^{MP}=\frac{1}{\sqrt{2}}\sqrt{\frac{N_{0}\pi}{\Theta k_{B}\beta}}e^{-\nicefrac{\Theta k_{B}\beta}{2}\left(N_{0}+1\right)}\textrm{erfi}\left(\sqrt{\frac{\Theta k_{B}}{2}\beta N_{0}}\right)\\
u_{N}^{MP}=\frac{\Theta k_{B}}{2}\left(1+N_{0}+\frac{1}{\Theta k_{B}}-\sqrt{\frac{2N_{0}}{\Theta k_{B}\beta\pi}}\frac{e^{N_{0}\left(\nicefrac{\Theta k_{B}\beta}{2}\right)}}{\textrm{erfi}\left[\sqrt{\frac{\Theta k_{B}}{2}\beta N_{0}}\right]}\right)\\
c_{N}^{MP}=\frac{k_{B}}{2}+k_{B}\sqrt{\frac{\Theta k_{B}\beta N_{0}}{2\pi}}\frac{e^{N_{0}\left(\nicefrac{\Theta k_{B}\beta}{2}\right)}}{\textrm{erfi}\left[\sqrt{\frac{\Theta k_{B}}{2}\beta N_{0}}\right]}\left(\frac{\Theta k_{B}\beta N_{0}}{2}-\frac{1}{2}-\sqrt{\frac{\Theta k_{B}\beta N_{0}}{2\pi}}\frac{e^{N_{0}\left(\nicefrac{\Theta k_{B}\beta}{2}\right)}}{\textrm{erfi}\left[\sqrt{\frac{\Theta k_{B}}{2}\beta N_{0}}\right]}\right).
\end{array}\label{mpthermfunct}
\end{equation}

\begin{figure}[tb]
\noindent \begin{centering}
\includegraphics{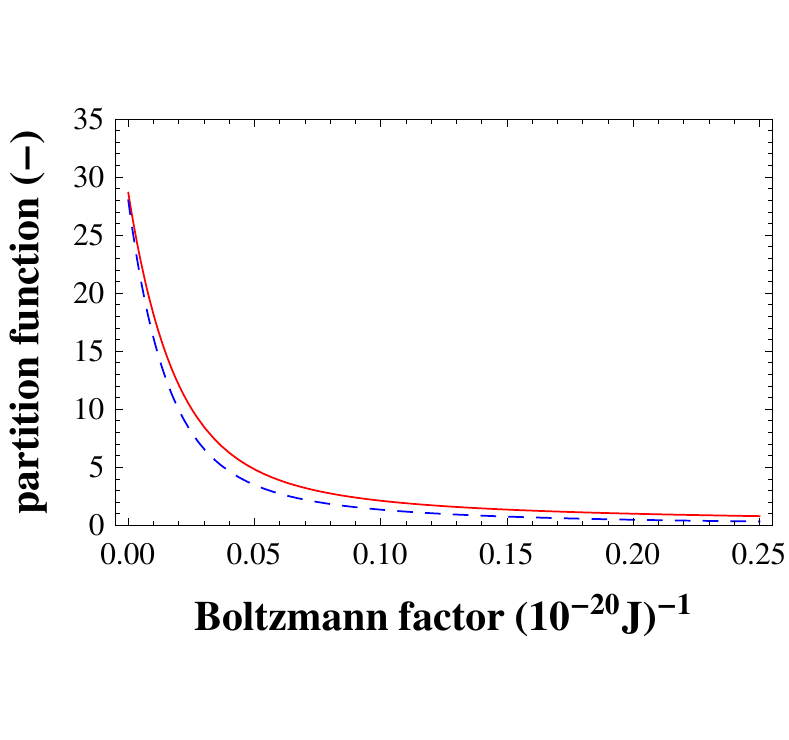} 
\par\end{centering}
\caption{\label{figcompartfunctgmpandmors.}Vibrational partition function
$z_{N}^{MP}$ (dashed line) and vibrational partition function $\tilde{z}^{GMP}$
(solid line) both as a function of $\beta$.}
\end{figure}

\begin{figure}[p]
\noindent \begin{centering}
\includegraphics{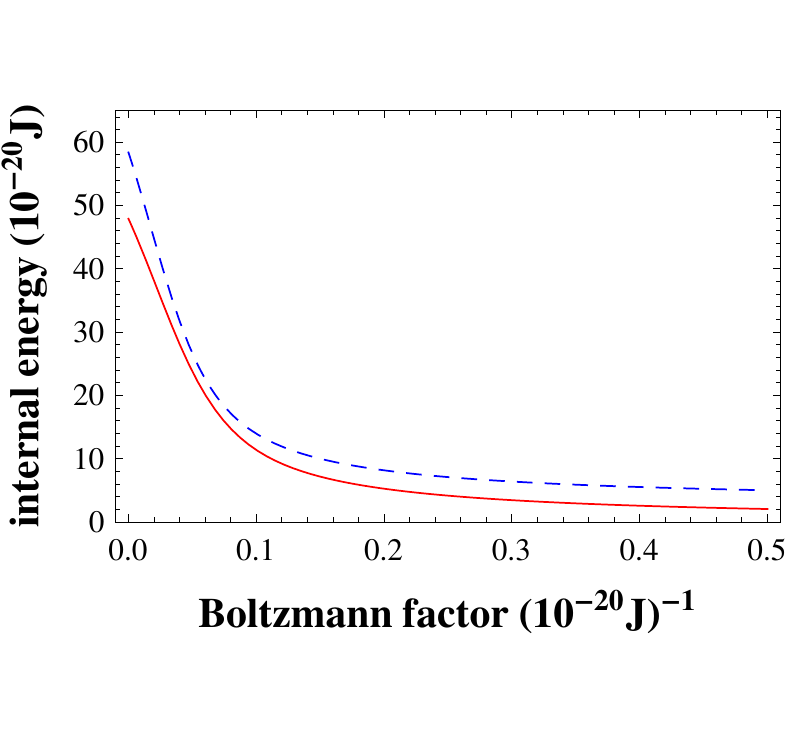} 
\par\end{centering}
\caption{\label{figcompintenergmpandmors}This graph represents the vibrational
internal energy $u_{N}^{MP}$ (dashed line) and vibrational internal
energy $u^{GMP}$ (solid line) both in terms of the $\beta$ parameter.}
\end{figure}

\begin{figure}[p]
\noindent \begin{centering}
\includegraphics{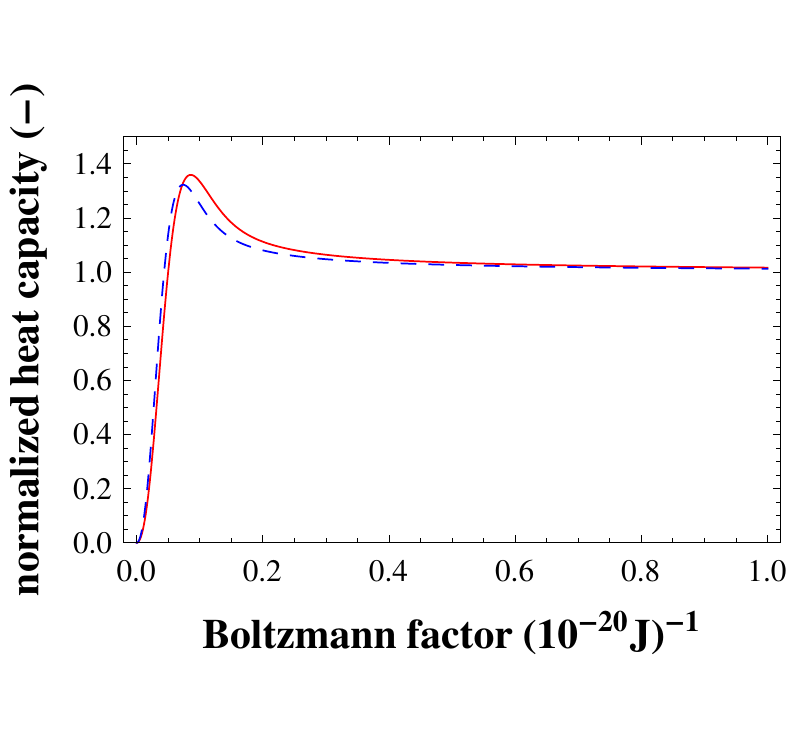} 
\par\end{centering}
\caption{\label{figcomspecheatgmpandmors.}Vibrational specific heat $\nicefrac{c_{N}^{MP}}{k_{B}}$
(dashed line) compared to vibrational specific heat $\nicefrac{c^{GMP}}{k_{B}}$
(solid line) as function of $\beta$.}
\end{figure}

As can be seen these functions also depend on the parameters of MP.
By following the previous work \cite{AngelovaFrank05} we take the
diatomic molecule $H\,Cl$ as an example for comparing the behavior
of the thermodynamic properties between GMP\textendash gas and MP\textendash gas.
It is important to note that in both gas models, GMP\textendash gas
(solid line) and MP\textendash gas (dashed line), these thermodynamic
properties depend explicitly on the $\textrm{erfi}(x)$. Figure \ref{figcompartfunctgmpandmors.}
shows the comparison between the individual partition functions in
the interval $\beta\in\left[0,\,1\right]$. Both curves have the same
behavior monotonous decreasing and reach zero asymptotically at the
same value of $\beta$. This is because both potentials contain almost
the same quantum states
. In Fig. \ref{figcompintenergmpandmors} the internal energies of
the two models also decay asymptotically for low temperature values,
however for $\beta\rightarrow0$, $u_{N}^{MP}\left(0\right)\rightarrow59.896\times10^{-20}\textrm{J}$
and $u^{GMP}\left(0\right)\rightarrow47.896\times10^{-20}\textrm{J}$.
This indicates that in general a system with a Morse interaction is
more energetic than a system with GMP interactions
. Finally, we note in Fig. \ref{figcomspecheatgmpandmors.} that the
specific heat for both models has the same intense effects of anharmonicities
in the region $\beta<0.168$. The maximum reached with the MP is shifted
to the left with respect to GMP, such that $T_{C}^{MP}\approx9815\textrm{K}$
and $T_{C}^{GMP}\approx8441\textrm{K}$. This means that a Morse molecular
interaction has the ability to energize many molecules and leave them
in the quantum states more energetic than GMP\textendash interaction,
which agrees with the behavior of the internal energy afore mentioned.
\pagebreak

\section{Concluding Remarks}

In the present paper, we have reaffirmed once again that the potential
derived from Deng and Fan models \cite{DengFan57}, as well as the
Manning\textendash Rosen type models \cite{ManningRosen33} are good
candidates to emulate the vibrational energy spectrum of the ground
state of several diatomic molecules. In order to get a better description
for heteronuclear diatomic molecules, in particular: $H\,F$,$H\,Cl$,$H\,Br$
and $H\,I$, we considered the algebraic model of GMP which turns
out to be a class of the Eckart potential due to its equivalences
with algebraic group $S\,O\left(2,1\right)$. Taking advantage that
the exact energy spectrum model for the EP is easy to operate, we
can obtain the individual partition function analytically and thus
derived the vibrational contribution to the thermodynamic functions
of these systems. Thermodynamic functions have already derived and
have the particularity that they depend on the interaction of the
model parameters (GMP) and hence the anharmonic effects are evident
in the region of high\textendash temperatures (between the dissociation
energy and vibrational characteristic energy). Proof of this is the
anomalous behavior in heat capacity \cite{MarzzaccoWaldman73}. However,
at $T\ll1$ we notice that the function $\nicefrac{c^{GMP}}{k_{B}}$
converges to the harmonic limit ($\left\llbracket \nicefrac{\left(\sqrt{B}-A\right)}{\alpha}\right\rrbracket \rightarrow\infty$),
that is because the individual vibrational partition function takes
into account a classical states density and therefore it continues
to fulfill the principle of energy equipartition. It is very important
to note, the algebraic model of $V_{GMP}$ keeps up a pseudo\textendash quantum
behavior. In order to obtain a complete description we need to introduce
a dynamic modelwhich includes the collective aspects of the molecules
that make up the gas \cite{GreinerNeiseStoecker95,Callen85}.

In order to verify the validity of the macroscopic system behavior
formed by molecules $H\,Cl$, we have compared some of the most important
thermodynamic functions with the results obtained by other authors
\cite{IkhdairFalaye13,OyewumiFalayeOnateOluwadareYahya14,Valenciaortega14}.
They used as potential interaction the Morse potential and the results
with GMP. At the same time we realised that $n_{max}\equiv\left\llbracket \nicefrac{\left(\sqrt{B}-A\right)}{\alpha}\right\rrbracket =28$
for GMP while $n_{max}\equiv N_{0}=28$ for MP. For that reason, the
form of the thermodynamic functions analyzed here do not differ to
a large degree. Although the exact solution to the energy spectrum
can be considered and its characteristic functions could be obtained,
the results show that the approach to high temperatures is good for
considering a classic state density. On the other hand, this study
may represent the first steps to characterize systems formed by lineal
polyatomic molecules with different chemical nature.

\section*{Acknowledgment}

This work was supported by SIP, COFAA, EDI\textendash IPN\textendash MÉXICO
and SNI\textendash CONACyT\textendash MÉXICO.


\begin{thebibliography}{Bibliography}
\bibitem{GreinerNeiseStoecker95}Greiner, W.; Neise, L.; Stöcker,
H. Thermodynamics and Statistical Mechanics; Springer\textendash Verlag
New York, Inc., \textbf{1995}; Chapter 8, pp 225\textendash 234. 

\bibitem{Callen85}Callen,H. B. Thermodynamics and An Introduction
to Thermostatistics; John Wiley and Sons, Inc., \textbf{1985}; Chapter
15, pp 329\textendash 332. 

\bibitem{McQuarrie76}McQuarrie, D. A. Statistical Mechanics; Harper
and Row Publisher, Inc., \textbf{1976}; Chapter 5, pp 91\textendash 108. 

\bibitem{Levine01}Levine, I. N. Quantum Chemistry; Pearson Education
S. A., \textbf{2001}; Chapter 13, pp 394\textendash 409.

\bibitem{Herzberg50}Herzberg, G. Molecular Spectra and Molecular
Structure, I Spectra of Diatomic Molecules; D. Van Nostrand Company,
Inc., \textbf{1950}; Appendix, p 501. 

\bibitem{Dong07}Dong, S\textendash H. Factorization Method in Quantum
Mechanics; Springer, \textbf{2007}; Chapter 6, pp 73\textendash 90.

\bibitem{DongLozadaYuJimenezRivera07}Dong, S\textendash H., Lozada\textendash Cassou,
M., Yu J., Jiménez\textendash Ángeles F. and Rivera A. L. \textit{Int.
J. Quantum Chem}. \textbf{2007}, \textit{107}, 366\textendash 371.

\bibitem{StaffordHoltPaulson63}Stafford F. E., Holt C. W. and Paulson
G. L. \textit{J. Chem. Educ}. \textbf{1963}, \textit{40}, 245\textendash 249.

\bibitem{DengFan57}Deng Z. H. and Fan Y. P. \textit{Shandong Univ.
J}. \textbf{1957}, \textit{7}, 162 (in Chinese).

\bibitem{CodrianskyCorderoSalam99}Codriansky S., Cordero P. and Salamó
S. \textit{J. Phys. A: Math. Gen}. \textbf{1999}, \textit{32}, 6287\textendash 6293.

\bibitem{AngelovaFrank05}Angelova M. and Frank A. \textit{Phys. Atomic
Nuclei} \textbf{2005}, \textit{68}, 1625\textendash 1633.

\bibitem{DelsolQuesneSmirnov98}Del Sol Mesa A., Quesne C. and Smirnov
Y. F. \textit{J. Phys. A: Math. Gen}. \textbf{1998}, \textit{31},
321\textendash 335.

\bibitem{Natanzon79}Natanzon G. A. \textit{Theor. Math. Phys}. \textbf{1979},
\textit{38}, 146\textendash 153.

\bibitem{CooperGinocchioKhare87}Cooper F., Ginocchio J. N. and Khare
A. \textit{Phys. Rev. D} \textbf{1987}, \textit{36}, 2458\textendash 2473.

\bibitem{Pathria96}Pathria R. K. Statistical Mechanics; Butterworth\textendash Heinemann,
\textbf{1996}; Chapter 6, pp 150\textendash 155.

\bibitem{ManningRosen33}Manning M. F. and Rosen N. \textit{Phys,
Rev}. \textbf{1933}, \textit{44}, 953\textendash 954.

\bibitem{MarzzaccoWaldman73}Marzzacco C. and Waldman M. J. Chem.
Educ. 1973, 50, 444\textendash 445.

\bibitem{IkhdairFalaye13}Ikhdair S. M. and Falaye B. J. \textit{Chem.
Phys}. \textbf{2013}, \textit{421}, 84\textendash 95.

\bibitem{OyewumiFalayeOnateOluwadareYahya14}Oyewumi K.J., Falaye
B.J., Onate C.A., Oluwadare O.J. and Yahya W.A. \textit{Mol. Phys}.
\textbf{2014}, \textit{112}, 127\textendash 141.

\bibitem{Valenciaortega14}Valencia-Ortega G. Contribución vibracional
anarmónica a la ecuación fundamental de un sistema de moléculas diatómicas.
Bachelor Thesis, Instituto Politécnico Nacional, Mexico City, March
2014 (in Spanish).
\end{thebibliography}
\end{document}